\documentclass{NAV}%

\usepackage{siunitx}

\usepackage{booktabs}
\usepackage{mathtools}

\usepackage{algorithm}
\usepackage{algpseudocode}

\newcommand{\R}{\mathbb R}
\newcommand\bfm{\mathbf m}
\newcommand\SA{\mathsf{SA}}
\DeclareMathOperator*{\argmax}{arg\,max}

\begin{document}

\title[Real-time rail vehicle localisation]{Real-time rail vehicle localisation using spatially resolved magnetic field measurements}

\author[Dieckow et al.]{Niklas Dieckow$^{1,2,4}$, Katharina Ostaszewski$^{2, 4}$, Philip Heinisch$^{2, 4}$, Henriette Struckmann$^{2, 4}$, Hendrik Ranocha$^{3, 4}$}

\address{\add{1}{Research Group on Human-Centric Machine Learning, Hamburg University of Technology (TUHH), Hamburg, Germany \email{niklas.dieckow@tuhh.de}}
\add{2}{PhySens Rail GmbH, Braunschweig, Germany}
\add{3}{Institute of Mathematics, Johannes Gutenberg University, Mainz, Germany}
\add{4}{Institute of Applied Numerical Science (IANW), Braunschweig, Germany}}

\begin{abstract}
This work presents two complementary real-time rail vehicle localization methods based on magnetic field measurements and a pre-recorded magnetic map. The first uses a particle filter reweighted via magnetic similarity, employing a heavy-tailed non-Gaussian kernel for enhanced stability. The second is a stateless sequence alignment technique that transforms real-time magnetic signals into the spatial domain and matches them to the map using a similarity measure. Experiments with operational train data show that the particle filter achieves track-selective, sub-5-meter accuracy over \SI{21.6}{\km}, though its performance degrades at low speeds and during cold starts. Accuracy tests were constrained by the GNSS-based reference system. In contrast, the alignment-based method excels in cold-start scenarios, localizing within \SI{30}{m} in \SI{92}{\percent} of tests (\SI{100}{\percent} using top-3 matches). A hybrid approach combines both methods—alignment-based initialization followed by particle filter tracking. Runtime analysis confirms real-time capability on consumer-grade hardware. The system delivers accurate, robust localization suitable for safety-critical rail applications.
\end{abstract}

\maketitle

\section{Introduction}\label{sec:introduction}

In the politically and societally driven transition towards sustainable mobility, rail transport plays a central role. To enhance the attractiveness and efficiency of rail transport, comprehensive structural and technological measures are indispensable. These include, among others, capacity expansion of existing infrastructure, improvements in the reliability and punctuality of rail services, and the implementation of autonomous train operations. Primary drivers for this include digitalisation and technological innovations, with vehicle-based real-time localisation playing a key role. Real-time localisation of trains enables optimised train scheduling and data-driven traffic management, such as on-demand services. Additionally, it serves as the foundation for train-oriented route safety (e.g., "Moving Block") and automated driving.

Current systems based on infrastructure-side markers or detectors (such as balises, axle counters) in combination with odometry systems do not fully enable this. Potential technologies for absolute localisation, such as satellite navigation (GNSS) and LiDAR (Light Detection and Ranging) for environmental perception, are being tested for safety-critical applications \citep{gnss_2015, lidar_2016, gnss_2018, gnss_2021, rev_loc_2022}. However, providing reliable positioning information in all relevant environments remains a significant challenge. The availability and precision of GNSS systems are compromised by shadowing and multipath propagation. In urban areas, signal quality is already notably reduced, leading to diminished performance and availability. In tunnels, the signal is completely blocked. For short periods, the quality and availability of GNSS can be enhanced through sensor fusion. Typically, inertial measurement units (IMUs) or odometry sensors are used for this purpose. However, these sensors only provide relative positioning information and can maintain high accuracy for only a few minutes.

Magnetic field-based localisation has emerged as a promising absolute localisation technology that can address the limitations of other approaches, such as availability issues \citep{heirich_2017, siebler_2018, zevrail_2024}. Magnetic field measurements have been explored in various applications, including aeroplane, indoor, and car localisation \citep{shockley_2014}. An analysis of the viability of magnetic field-based localisation for train systems, along with its associated challenges, was conducted by \citet{heirich_2017}. More recently, \citet{siebler_2018} introduced a similar method for along-track train localisation using a particle filter that relies solely on magnetic field measurements. The authors later enhanced this method with a Kalman filter, enabling simultaneous train localisation and magnetometer calibration \citep{siebler_2023}. Additionally, a similar approach that utilises along-track variations in magnetic permeability has been investigated by \citet{spindler_2016, spindler_2018, Kroeper_2020}.

Besides contributing to the existing results on warm start localisation with a particle filter, we also present empirical results on a cold start localization task using two kinds of methods: the particle filter approach, and an approach based on sequence alignment in the spatial domain, as proposed by \citet{heirich_2017}. Finally, we propose a combination of the two methods that leverages the strengths of each of the methods, and discuss its real-time viability.

\section{Methodology}\label{sec:methodology}
For the localisation, we employed two different approaches. The first is a particle filter, similar to the one described in the work of \citet{siebler_2018}, while the second relies on a subsequence search in the spatial domain.

\subsection{Particle Filter}
The first approach maintains a large collection of independent particles $\{p_i\}_{i=1}^N$, each of which have a state ${(x_i, v_i)^\top \in \R^2}$ consisting of along-track position and velocity, as well as an associated weight $w_i \in [0,1]$ such that $\sum_{i=1}^N w_i = 1$. At every update step, the state of each particle is updated using a noised movement model of the form
\begin{equation}
    \begin{pmatrix}
        x^{(t+1)} \\
        v^{(t+1)}
    \end{pmatrix}
    \sim
    \mathcal N\left(\begin{pmatrix}
        1 & \Delta t \\
        0 & 1
    \end{pmatrix}
    \begin{pmatrix}
        x^{(t)} \\
        v^{(t)}
    \end{pmatrix}, \mathbf Q\right),
\end{equation}
where $\Delta t > 0$ is the reciprocal of the filter's update frequency and $\mathbf Q$ the noise covariance matrix, defined by
\begin{equation}
    \mathbf Q = q \begin{pmatrix}
        \Delta t^3 / 3 & \Delta t^2 / 2 \\
        \Delta t^2 / 2 & \Delta t
    \end{pmatrix},
\end{equation}
following the noise model proposed by \citet{crouse_2023}.

Using a pre-recorded magnetic map, the magnetic field measurements at each particle's location can be retrieved and compared to the measurements which are received in real time by the sensor. Using a weighting function, or kernel, $K \colon \R^3 \times \R^3 \to \R$, particles whose associated magnetic field measurements are closer to the observed ones are weighted higher, and those further away are weighted down,
\begin{equation}
    \widetilde w_i^{(t+1)} = w_i^{(t)} \cdot K\left(\mathbf m_S(x_i^{(t)}), \mathbf m_T^{(t)}\right), \quad w_i^{(t+1)} = \frac{\widetilde w_i^{(t+1)}}{\sum_{j=1}^N \widetilde w_j^{(t+1)}}.
\end{equation}
Typically, a kernel derived from the Gaussian distribution is used for particle re-weighting. However, for practical reasons, we instead utilised the kernel
\begin{equation}\label{eq:mod_kernel}
    K(\mathbf x, \mathbf y) = \frac{1}{1 + \lVert \mathbf x - \mathbf y \rVert},
\end{equation}
which provided greater stability in our experiments, compared to the Gaussian kernel, likely due to its heavier tails. The state of the particle filter consisted of the train position and velocity, which in turn means that the particle filter provides a velocity estimate as well. %

\subsection{Alignment-Based Approach}
The second approach is based on time series similarity estimation. The key idea is to first transform the most recent chunk of the measured, time-based, magnetic field signal into a position-based signal, followed by a comparison with all chunks of equal length in a pre-recorded magnetic map. In other words, the problem is transformed into a sequence of time sub-series search problems. These can be solved by computing distances in a sliding window-manner. Possible distance metrics include the Euclidean distance and Dynamic Time Warping \citep{dtw}. This method is further described in the following sections.

The two methods differ in terms of statefulness. The particle filter maintains, for each particle, a position and velocity, and updates these values after each incoming measurement. In contrast, the sequence alignment-based approach is stateless: As new data comes in, the sub-series search starts anew, without incorporating any prior position information. Because of this, we expect the particle filter to be better suited for the task of continuous localisation, as the position of the train at time $t$ places a strong prior on the position at time $t+1$. This could however be a disadvantage for initially finding the train's location -- a task which we term \emph{cold start localisation} -- as the particle filter might latch onto a poor prior.
It should be noted that the sequence alignment method can also be modified to yield a stateful method by explicitly restricting the search region, similar to what has been done by \citet[Section 9]{heirich_2017}.

Both methods rely on the existence of a magnetic map, which is a function $\bfm \colon [-90, 90] \times [-180,180] \to \R^3$ mapping from a coordinate given in terms of latitude and longitude to the corresponding magnetic field measurement. This map must be created in advance by traversing the entire track once by train and recording the magnetic field measurements. 

\section{Mathematical Preliminaries}\label{sec:prelims}
In the following, let $t \in \R_{\geq 0}$ denote the \textbf{time} and $x(t) \in \R_{\geq 0}$ the \textbf{along-track position} at time $t$.
In order to perform similarity computations between the live magnetic field measurements $\bfm_T(t)$, which are available in time coordinates, and a pre-recorded magnetic map $\bfm_S(x)$, which is available in spatial coordinates, a way of transitioning from temporal to spatial coordinates is necessary. This is achieved by a non-linear coordinate transformation. We first consider the idealised continuous perspective, before looking at the actual implementation.

We start with the transformation from space to time. Assuming that a velocity function $v(t) = x'(t)$ and the starting position $x(0)$ are known, we can simply integrate to obtain the along-track position $x(t)$. This readily yields a magnetic map in time coordinates via $\bfm_T(t) = \bfm_S(x(t))$.

Conversely, to obtain a time-to-space transformation, we can invert $x(t)$, yielding $t(x)$. This inverse exists, if $v(t) > 0$ for all $t$ in the considered time horizon, which corresponds to the restriction that the train always moves, and only in one direction.
In practice however, the assumption of $v$'s positivity can fail, as the train may stand still or drive backward. This needs to be detected and handled separately.

We note that it is possible to cast the problem of finding $t(x)$ as an ordinary differential equation,
\begin{equation}
    t'(x) = \frac{1}{v(t(x))},
\end{equation}
which follows directly from the inverse function theorem. An analytic solution of this ODE leads us back to the equation $t = x^{-1}$, the inverse function of $x$. Nonetheless, one advantage of the ODE formulation is that we may apply different numerical solution methods in practice, such as the Euler method, rather than purely relying on inverse computation methods.

\subsection{Preprocessing}\label{sec:preprocessing}
In the following, we describe the two preprocessing steps of velocity estimation and spatial transformation in more detail.
\subsubsection{Velocity Estimation}
To obtain a velocity signal, different methods are available. During our experiments, we made use of velocity data obtained from an inertial measurement unit (IMU), but it is also possible to leverage the signal shift between two magnetic sensors placed very closely next to each other.
In the latter method, we rely on two streams of magnetic measurement data $\bfm_1, \bfm_2$, respectively coming from two magnetometers mounted at a fixed distance $d$ apart from each other on the same side of the train. Due to this set-up, we know that $\bfm_2(t) = \bfm_1(t - \tau(t))$ for some time-dependent time delay $\tau(t) > 0$. The problem of estimating this delay from the signals is known as \emph{time delay estimation} and can be solved by performing a cross-correlation on the last $N$ samples of both signals \citep{Azaria_1984}, where $N$ is a hyperparameter. The velocity can then be approximated by $\tilde v(t) = \frac{\tau(t)}{d}$ \citep{zevrail_2021}.

\subsubsection{Domain Transformation}
For the time-to-space transformation of the magnetic field signal, we require the inverse of $x$. Assuming that the time points are equidistantly sampled, this can be done via reverse interpolation.
To this end, denote by ${\underline t = (t_1, \dots, t_n) \in \R^n}$ the sequence of discrete equidistant time points such that ${t_{i+1} = t_i + \Delta t}$ for all $1 \leq i \leq n-1$ and for some $\Delta t > 0$. From this we can define discretisations $\underline v = (v(t_i))_{i=1}^n \in \R^n$ and $\underline x = (x(t_i))_{i=1}^n \in \R^{n}$. The latter can be directly obtained as a cumulative sum of $\underline v$ multiplied by $\Delta t$ and added to the initial value, i.e.,
\begin{align}
    \underline x =
    \begin{pmatrix}
        x(0) + \Delta t v_1 \\
        x(0) + \Delta t (v_1 + v_2) \\
        \vdots \\
        x(0) + \Delta t(\sum_{i=1}^n v_i)
    \end{pmatrix}.
\end{align}
Due to the shared dimension, we can map $x_i \mapsto t_i$ for each $1 \leq i \leq n$, and obtain an approximation to $\bfm_S$ by interpolating:
\begin{equation}
    \widetilde{\bfm}_S(x) \coloneqq \bfm_T(f(x)),
\end{equation}
where $f \coloneqq \mathsf{interp}(\underline x, \underline t)$ is the result of some kind of interpolation method, ${\mathsf{interp} \colon \R^n \times \R^n \to C(\R)}$, that guarantees equality at the grid points. In our experiments, linear interpolation was used.
Finally, $\widetilde{\bfm}_S$ is sampled at a frequency of $f_S = 1 / \Delta x$, which coincides with the sampling frequency of the pre-recorded magnetic map, yielding a discrete spatially-resolved magnetic field signal $\widetilde{\underline\bfm}_S \in \R^m$ that can be used for comparison.

\subsection{Similarity of spatial signals}
The obtained signal $\widetilde{\underline\bfm}_S$ can be aligned with the pre-recorded magnetic map $\underline\bfm_S$ using any sequence alignment method. Denote any such method by $\SA \colon \R^n \times \R^m \to \{1,\dots,n-m\}$, where we assume $n \geq m$. Then ${i = \SA(\mathbf a, \mathbf b)}$ is an index such that $\mathbf a_{i:i+m}$ and $\mathbf b$ are a best match in some arbitrary sense.

Any method $d \colon \R^m \times \R^m \to \R_{\geq 0}$ for computing some distance between two same-length sequences can be turned into a sequence alignment method via a sliding window approach,
\begin{equation}
    \SA_d(\mathbf a, \mathbf b) = \argmax_{i \in \{1,\dots,n-m\}} d(\mathbf a_{i:i+m}, \mathbf b).
\end{equation}

Examples of this would include the Euclidean distance or the Dynamic Time Warping distance \citep{dtw}, defined as the minimal Euclidean distance across all monotone deformations of the original sequences. In our experiments, we found the latter to produce the best results.

\section{Experimental Set-Up}\label{sec:experimental-set-up}
\begin{figure}
    \centering
    \includegraphics[width=\linewidth]{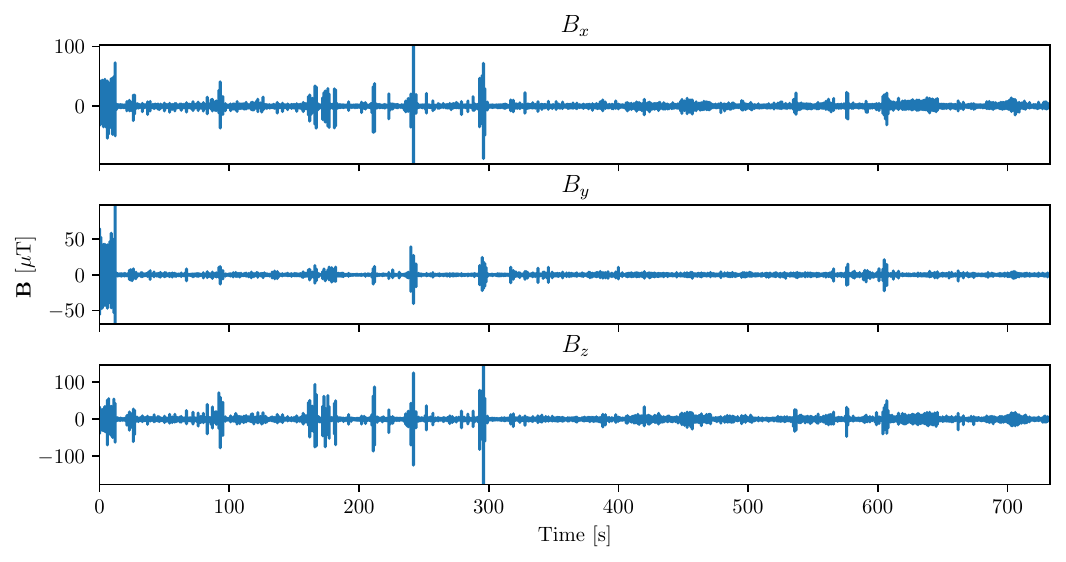}
    \caption{Magnetometer data of a \SI{21.6}{\km} long portion of the RAILab data that was used in the experiments.}
    \label{fig:bav_20km_data}
\end{figure}

Two different sets of rail vehicles were used to gather magnetic field measurements along the tracks. To ensure the validity of the data, all measurements were performed during regular service without special preparations or operational changes.
ODOMAG sensor units \citep{zevrail_2021, zevrail_2024} were mounted beneath the vehicles at different positions depending on the mechanical design and space availability. Each sensor unit consists of four sets of tri-axial magnetometers with the necessary control and processing electronics housed together in a rectangular V4A stainless-steel tube (approx. \SI{4}{\cm} $\times$ \SI{4}{\cm} $\times$ \SI{70}{\cm}). Power and data transmission are handled via a single 802.3af Ethernet connection. The sensor units provide a synchronised raw 24-bit output for each magnetometer with a sampling rate of \SI{80}{\kHz}. 
\begin{enumerate} 
\item \textbf{RegionAlps:}
Multiple sensor units were mounted beneath the locomotive of a RBDe 560 DOMINO trainset operated by Swiss “Region Alps”, directly above the rail next to one of the bogies with a distance between the sensor and rail of approx. \SI{24}{cm}.

Velocity reference was provided redundantly by a combined Galileo and GPS GNSS receiver unit on the inside next to the door of the driver compartment and a wheel impulse generator attached to a non-driven wheel set.
\item \textbf{RAILab:} 
A magnetic sensor unit was mounted below a non-powered track maintenance car of the German Deutsche Bahn (DB) directly above the rail with a distance between the sensor and rail of approx. \SI{50}{\cm}.

Two redundant systems were used as reference for position and velocity. One system consists of a combined Galileo and GPS GNSS receiver unit mounted on the outside of the rail car directly adjacent to the sensor. A second reference was provided by the combined GNSS and wheel impulse generator system used for track localisation purposes during maintenance.  
\end{enumerate}

\section{Evaluation}\label{sec:evaluation}
We evaluated the methods on two different tasks. The first is \textbf{continuous localisation}, in which the initial position is known and the localisation method is active during the entire train ride. It tests both the long-term stability and the real-time capabilities of the method. The main evaluation criterion for this task is the localisation error over time. This is computed using the Haversine distance between the estimated position and the true position. Because our implementation of the sequence-based method is stateless, it is not suitable for continuous localisation and we only evaluated the particle filter on this task.
The second task is \textbf{cold start localisation}, in which the initial position is unknown and the method is only active until the correct position has been ``found,'' in a precise sense that will be defined shortly. This task is relevant because it may be desirable to switch to a different algorithm once the correct position has been found and the set of possible on-track positions can be narrowed down.

\subsection{Continuous localisation}\label{sec:continuous_loc}
For the warm start localisation task, we took the longest section that did not contain any stops, which turned out to be \SI{21.6}{\km} in length. The magnetic field data along this section is displayed in Figure \ref{fig:bav_20km_data}. The reason we want to avoid stopping is due to the particle filter's tendency to diverge during passages with low velocities, which we loosely define as velocities less than \SI{10}{\m/\s}.

\begin{figure}[htb]
    \centering
    \includegraphics[width=1\linewidth]{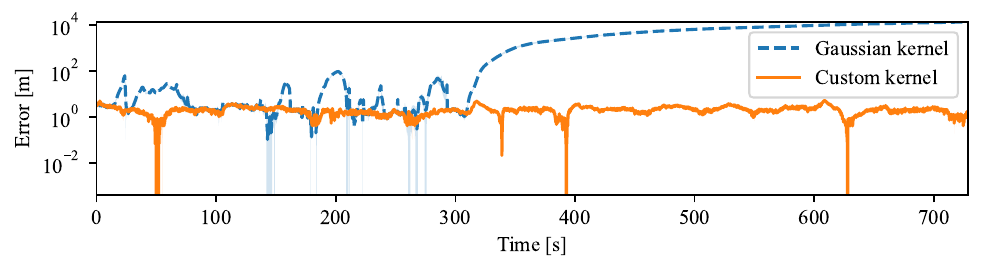}
    \caption{Error over time of the continuous, warm-start localisation using a particle filter on the \SI{21.6}{\km} section of the RAILab data. The blue curve shows the results using a Gaussian kernel for weight updating, while the orange curve corresponds to the modified kernel \eqref{eq:mod_kernel}. The results were averaged across 10 runs and the 3$\sigma$-interval is shown as a shaded region, although mostly invisible due to the low variance in the results. All runs used $N = \SI{10000}{}$ particles and noise coefficients $q = 0.53$ and $q_\mathrm{gauss} = 0.67$, respectively. For the Gaussian kernel, the scale coefficient was set to $\sigma = 10$, which provided the most stable results.}
    \label{fig:continuous_localisation}
\end{figure}

Figure \ref{fig:continuous_localisation} displays the warm-start localisation results on this section of the track, once for the Gaussian kernel and once for the modified kernel \eqref{eq:mod_kernel}. Evidently, the Gaussian kernel, despite using the best-performing scale value of $\sigma = 10$, diverges about halfway into the track. Among all runs with the modified kernel, the mean position error was \SI{2.07}{\m} with a maximum error of \SI{5.3}{\m}. We note that these results are distorted by the position error inherent in the magnetic map. As a consequence, the true position error may be even lower than what our results indicate.

\subsection{Cold Start}\label{sec:cold_start}
With the cold start experiments, we investigated the ability and required track length of the methods to localise the train, given that no prior knowledge about the train's location is available, apart from the track bounds.

For the purpose of these experiments, we loosely define that a method has \emph{found the track} or simply \emph{converged}, once a localisation error of less than \SI{25}{\m} has been reached, which corresponds to the typical length of a train car. This is also well within the range of particle filter initialisation schemes in previous works, such as \citet{siebler_2018}, where the authors initially distribute the particles within \SI{50}{\m} of the true position.

\subsubsection{Particle Filter}
Using the track from Figure \ref{fig:bav_20km_data}, we uniformly distributed 100,000 particles across the entire track and set the noise parameter to $q = 0.2$. However, the filter did not converge, even for different choices of the two parameters.

For a second experiment, we used data from a roughly \SI{66}{\km} long track connecting Saint-Maurice and Leuk in Switzerland. We used the intermediate stops as different starting points for the cold start localisation. Because the particle filter had difficulties converging at slow speeds, we only began running it once the train velocity had reached \SI{10}{\m/\s}, in order to avoid the convergence to a bad prior.

\begin{figure}
    \centering
    \includegraphics[width=1\linewidth]{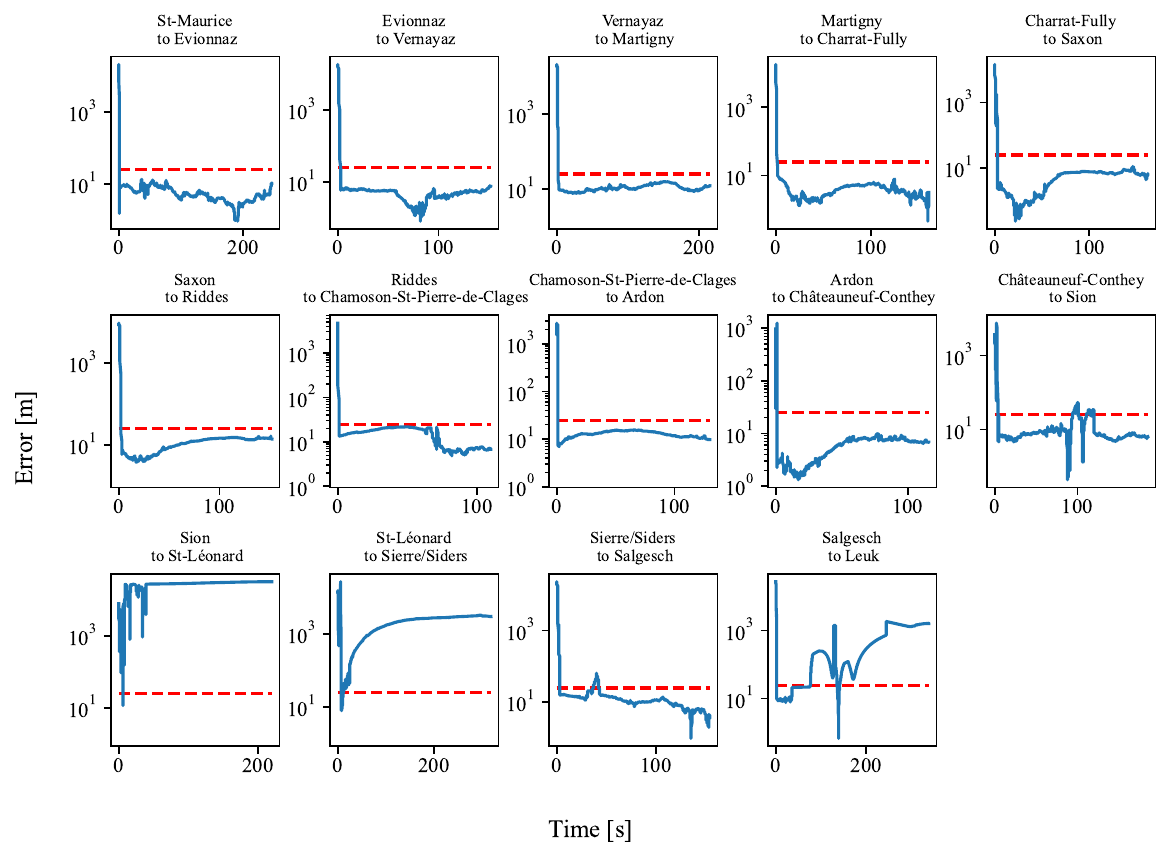}
    \caption{A time vs. error plot of the results from the cold start localisation experiment on the Switzerland data. Note the logarithmic y-axis. The horizontal dashed (red) lines demarcate the \SI{25}{\m} error threshold.}
    \label{fig:pf_cold_start_swiss}
\end{figure}

As before, the particle filter was initialised with 100,000 uniformly distributed particles across the entire \SI{66}{\km} track and the noise coefficient was set to $q = 0.2$. We then recorded the localisation error over time. The results are shown in Figure \ref{fig:pf_cold_start_swiss}. We observe that a sustained sub-\SI{25}{\m} localisation was successful in 11 out of the 14 sections, although the particle filter briefly exits the \SI{25}{\m} error zone in two of those sections. In the remaining three sections, localisation was either initially successful but diverged over time, or never succeeded in the first place. The average time until localisation was \SI{3.1}{\second} ($\sigma$ = \SI{1.7}{\s}). In terms of distance, localisation occurred after \SI{37.3}{\m} on average ($\sigma$ = \SI{24.8}{\m}). Excluding the three diverging sections, this lowers to a mean of \SI{27.9}{\m} and a standard deviation of \SI{10.9}{\m}.

Finally, we mention that two of the three diverging sections (Sion $\rightarrow$ St-Léonard and St-Léonard $\rightarrow$ Sierre/Siders) do converge when starting the particle filter at velocities of \SI{15}{\m/\s} and \SI{25}{\m/\s}, respectively. For Salgesch $\rightarrow$ Leuk, this was not the case, even for higher starting velocities.

\subsubsection{Alignment-Based Method}
As the sequence alignment-based method is stateless, we investigated how large (i.e., how many meters long) the subsequence has to be in order for a location with an error below \SI{25}{\m} to be within the set of $k$ most likely predictions, for $k \in \{1,2,3,4,5\}$. Top-$k$ predictions with $k>1$ can be useful, as one can weed out the incorrect ones by, for instance, starting a particle filter at each location contained in the top-$k$ result, and drop those that do not converge. This is further discussed in Section \ref{sec:unification}.

We considered discrete subsequence lengths $\ell$ of $1,2,\dots,\SI{200}{\m}$. We call $L(k)$ the minimum value of $\ell$ required in order for any of the $k$ best matches to have a Haversine error of at most \SI{25}{\m}.

For the \SI{21.6}{\km} long portion from the RAILab data, the top-1 prediction error was 
\SI{190.1}{\m} for the worst choices of $\ell$. We note that while these worst choices correspond to the shortest lengths (between \SI{1}{\m} and \SI{3}{\m} specifically), the error may surprisingly increase with increasing $\ell$, although this behaviour appears to stabilize for large enough $\ell$. Table \ref{tab:align_coldstart_bav_results} reports the different values of $L(k)$. We see that it is necessary to look beyond the top-1 match, as even a subsequence length of \SI{200}{\m} was not sufficient to guarantee localization.

\begin{table}[]
    \caption{Shortest required subsequence length required to achieve sub-\SI{25}{\m} error localisation, as a function of the number of best $k$ matches considered. These results are based on the \SI{21.6}{\km} portion from the RAILab dataset.}
    \label{tab:align_coldstart_bav_results}
    \centering
    \begin{tabular}{cccccc}
        \toprule
        $k$ & 1 & 2 & 3 & 4 & 5 \\
        \midrule
        $L(k)$ [m] & >200 & 109 & 100 & 74 & 34 \\
        \bottomrule
    \end{tabular}
\end{table}

For the Swiss data, we performed the experiment for each of the different stops, 14 in total. The results are listed in Table \ref{tab:align_coldstart_results}. Compared to the particle filter, the mean and maximum distance until localisation is considerably lower for all values of $k$.
Furthermore, we can see that there is a benefit to including the second-best match, as it eliminates the problem of not finding one of the tracks. Additionally, considering the third-best match reduces the maximum subsequence length by another \SI{10}{\m}, as well as the mean and standard deviation. Beyond that, however, there is no significant benefit to considering further matches.

Concluding this section, we summarise that, while the particle filter is capable of performing cold start localisation within just \SI{30}{\m}, it is challenged by slow speeds, and sometimes fails entirely, evidenced by the \SI{78.5}{\percent} success rate during the experiments with the RegionAlps data, and the failure to converge at all when tested on the RAILab data. Since the sequence alignment-based method works with spatial data, slow speeds are not an issue, provided that the conversion from the temporal to the spatial domain is accurate. We observed higher success rates (\SI{92.2}{\percent} for top-1 and \SI{100}{\percent} for top-$k$ with $k \geq 2$) and shorter distances until convergence, almost 2.5 times less than the particle filter, for $k=3$.

\begin{table}[]
    \caption{Results of the cold start experiments for the sequence alignment-based method on the RegionAlps data. Reductions (mean, standard deviation, min, max) are applied across the different stops (14 in total). The last column is the number of stops (out of 14) for which not even a \SI{100}{\m} subsequence length was sufficient. For a comparison with Table \ref{tab:align_coldstart_bav_results}, the "maximum" column may be seen as an equivalent to $L(k)$.}
    \label{tab:align_coldstart_results}
    \centering
    \begin{tabular}{cccccc}
        \toprule
        $k$ & Mean [m] & Std. deviation [m] & Minimum [m] & Maximum [m] & \# Not found \\
        \midrule
        $1$ & $19.0$ & $19.74$ & $1$ & $80$ & 1 \\
        $2$ & $14.64$ & $17.76$ & $1$ & $76$ & 0 \\
        $3$ & $11.5$ & $15.83$ & $1$ & $66$ & 0 \\
        $4$ & $10.79$ & $15.44$ & $1$ & $65$ & 0 \\
        $5$ & $10.79$ & $15.44$ & $1$ & $65$ & 0 \\
        \bottomrule
    \end{tabular}
\end{table}

\section{Unification and Real-Time Capabilities}\label{sec:unification}
In this section, we outline one possibility of combining the particle filter and the alignment-based method and discuss its real-time efficacy by drawing upon performance data of our experiments, as well as the computational complexity of the algorithms.

Based on the considerations and experiments in the previous section, the procedure outlined in Algorithm \ref{algo:localisation_unified} combines the strengths of both approaches well. While it is described as an offline algorithm, where the full magnetic signal $\mathbf m_T$ is available, it works the same in an online scenario. The pseudocode utilises the following subroutines which we describe as text only:
\begin{itemize}
    \item \textsc{Spacify}($\mathbf m_T$): Transforms the time signal into a corresponding space signal $\widetilde{\underline\bfm}_S$ as described in Section \ref{sec:preprocessing},
    \item $\textsc{AlignTop3}(\mathbf x, \mathbf y)$: attempts to locate $\mathbf y$ in $\mathbf x$ via some sequence alignment method and returns the best three matches,
    \item $\textsc{GetPF}(x_1,x_2,x_3,\ell,\tau)$: Initialises three particle filters at the respective points $(x_1,x_2,x_3)$, runs them in parallel until all but one have diverged (that is, the particle variance has exceeded the threshold $\tau$) and returns the remaining particle filter; if after $\ell$ updates, less than two filters have diverged, it is checked whether the converged filters coincide: if so, only one of them is returned. If not, or if all three filters have diverged, \texttt{Error} is returned. Along with the possible outputs described above, the current location index $x$ is returned as well.
\end{itemize}

\begin{algorithm}
    \caption{Offline localisation}\label{algo:localisation_unified}
    \begin{algorithmic}[1]
        \Require magnetic map $\bfm_S$, discrete magnetic signal $\underline\bfm_T$, lookback range $n$, max. PF updates $\ell$, divergence threshold $\tau$
        \Procedure{Localise}{$\bfm_S, \underline\bfm_T, s, e, n$}
            \State $x \gets n$
            \State $\widetilde{\underline\bfm}_S \gets \textsc{Spacify}(\underline\bfm_T)$
            \State $x_1,x_2,x_3 \gets \textsc{AlignTop3}(\bfm_S,\widetilde{\underline\bfm}_S\texttt{[x-n:x]})$
            \State $\mathcal P, x \gets \textsc{GetPF}(x_1,x_2,x_3,\ell,\tau)$
            \If{$\mathcal P = \texttt{Error}$}
                \State \textbf{go to} line 4
            \Else
                \State run $\mathcal P$ until divergence \textbf{then go to} line 4
            \EndIf
        \EndProcedure
    \end{algorithmic}
\end{algorithm}

Apart from the choice of alignment method, the algorithm has the following hyperparameters: the lookback range $n \in \mathbb N$, particle variance threshold $\tau \in (0,\infty)$ and the length $\ell \in \mathbb N$ of the ``burn phase'' of the particle filters. As a guideline for choosing these parameters, we recommend looking at the cases where $\textsc{GetPF}$ returns \texttt{Error}: If it happens because there are at least two convergent filters at different positions, consider increasing $n$, decreasing $\tau$ or increasing $\ell$. If all filters diverge, consider increasing $\tau$ or checking the data.

\subsection{Computational Complexity and Real-Time Capabilities}
Based on Algorithm \ref{algo:localisation_unified}, the localisation begins with a subsequence search over the entire reference sequence.
The time complexity of dynamic time warping is $\mathcal O(NM)$, where $N$ is the length of the reference sequence and $M$ the length of the query sequence. As a sequence-alignment method, when naively implemented, the complexity would involve $N^2$, but it is possible to maintain the original complexity when computing sequence alignment, see also \citet[Section 7.2.3]{music_processing}. Based on the experimental results in Table \ref{tab:align_coldstart_results}, a query sequence length of \SI{100}{\m} is more than sufficient for successful cold start performance. On consumer-grade hardware\footnote{Computations were performed on a laptop equipped with an AMD Ryzen 7 PRO 7840U CPU and 32GB LPDDR5 RAM (6400 MT/s).}, using the DTW-based sequence matching implementation provided by the Python package \emph{DTAIDistance} \citep{dtai_distance}, cold start localisation with \SI{100}{\m} query length over a \SI{66}{\km} reference track took \SI{4.28}{\second} on average. In practice, a rough estimate of the location is usually known, allowing these computation times to be kept to a minimum.

Particle filters allow for an extremely efficient implementation, mainly by means of parallelisation, as all particles are independent from one another. For Algorithm \ref{algo:localisation_unified} to work in real time, the particle filter updates should be faster than the train speed. This is because the initial step, \textsc{AlignTop3}, might take a few seconds as discussed before, during which the position of the train will have changed and additional time-resolved data will have been collected. In order to catch up to the real-time train position, the particle filter needs to process this buffered data quickly. For reference, using the same specs as above, a particle filter with 100,000 particles and 10 updates per second was able to process real-time data sampled at \SI{2000}{\Hz} roughly 12 times faster on average.

\section{Conclusion and Future Work}\label{sec:conclusion}
In this work, we investigated two approaches for the task of rail vehicle localisation using magnetic field measurements. While both methods rely on a pre-recorded magnetic map, their use of this map for the live localisation task differs. The first method is a particle filter, which has been proposed and investigated for this problem in previous works. Based on our experiments, it performs well at following the train's position when the latter is known from the start. It is also easily parallelisable. The downsides of this method are its nondeterministic nature, making it impossible to rule out an eventual divergence of the filter, as well as its unreliability at low velocities.
The second method relies on sequence alignment after a transformation into spatial coordinates. This method is deterministic and performs particularly well at initially localising the train using less than \SI{100}{\m} of track length, even when prior information about the train's position spans several tens of kilometres. While the particle filter was capable of performing the same task, the alignment-based method did so with higher accuracy and faster convergence.

Finally, we proposed a combination of both approaches that uses the alignment-based method to find an initialisation point from which, once the train speed is high enough, a small number of particle filters can be started. We discussed the computational complexity of this approach, which, along with the provided computation times of our experiments, provide strong evidence for the real-time viability of the approach.

Due to the nature of the particle filter, it can also easily be integrated with hardware accelerometers to increase reliability at low velocities and prevent eventual divergence. This kind of sensor fusion approach will be one of the primary aspects of future work to integrate the alignment-based and filter-based methods with an accelerometer or even and additional gyroscope for full inertial measurement.

\section*{Acknowledgment}

The authors thank RegionAlps SA and Enotrac AG for their valuable support and collaboration in this research. We are especially grateful for the opportunity to integrate our sensor system into one of RegionAlps vehicles and conduct tests under real operational conditions.

This research was partially funded by the Federal Ministry of Transport and Digital Infrastructure (Bundesministerium für Digitales und Verkehr, BMDV) through the mFUND program. We gratefully acknowledge their support under the FKZ 01F1128.

\section*{Competing Interests}
The authors declare no competing interests.

\bibliography{bibliography}

\end{document}